\documentclass[twocolumn,amssymb,prl,superscriptaddress,floatfix,showpacs]{revtex4-1}

\usepackage{graphicx}
\usepackage{amsmath, amssymb}

\begin{document}

\newcommand{\be}{\begin{equation}}
\newcommand{\ee}{\end{equation}}

\newcommand{\PT}{PT }
\newcommand{\sech}{{\rm sech}}

\newcommand{\placeFigNormalAndUs}{\begin{figure}[b]
\centering
\includegraphics[width=7cm]{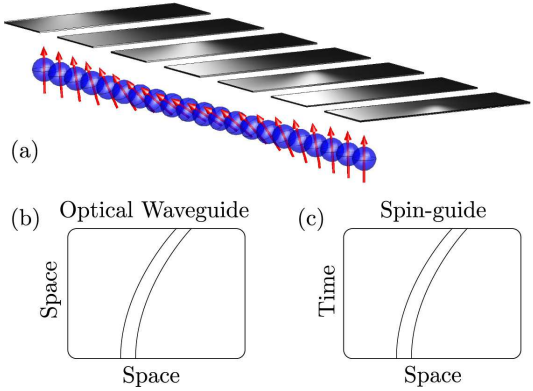}\\
\caption{(a) Schematic of a spin-guide: a one-dimensional line of
  spins is positioned below a gate array.  The gate potentials are
  varied, breaking the translation symmetry of the chain, to define a
  spin-guide capable of carrying an excitation. The size of the gates
  is expected to be much larger than the inter-spin spacing so that
  individual control of the spins is not possible.  (b) A conventional
  waveguide is defined by a local change in the refractive index of a
  medium.  This can be thought of as defining a two-dimensional
  pathway. (c) A spin-guide is defined by a 1+1 dimensional variation
  in the spin properties, which mimics the refractive index profile of
  an optical waveguide.}
\label{fig:normalAndUs}
\end{figure}}

\newcommand{\placeFigMoving}{\begin{figure}[t!]
\centering
\hspace{-0.7cm}\includegraphics[height=2cm]{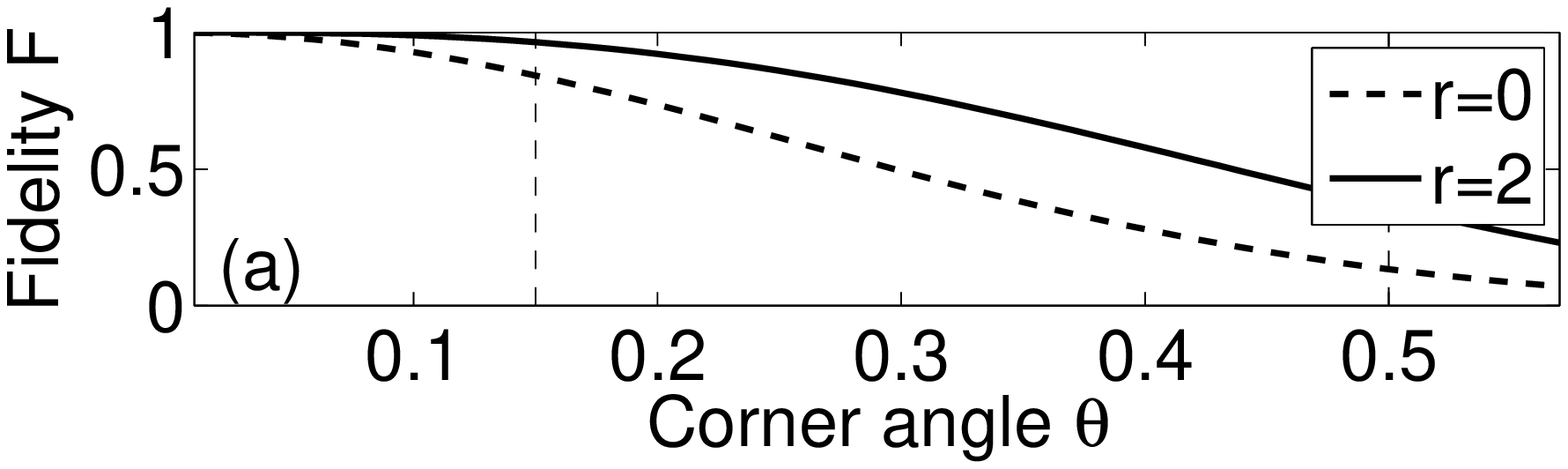}\\
\includegraphics[width=7cm]{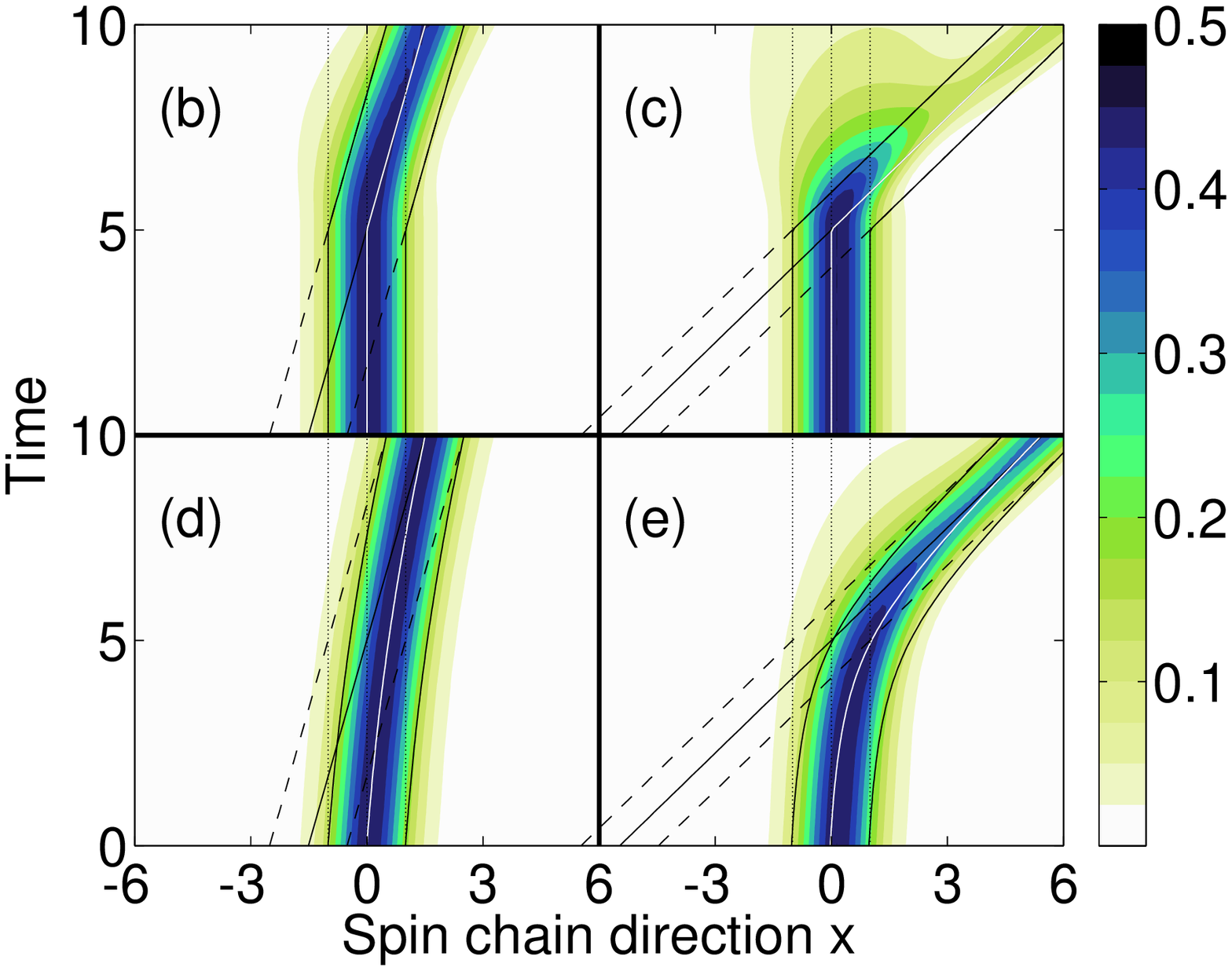}
\caption{(a) Fidelity as a function of corner angle $\theta$, for two
  values of $r$: $r=0$ (discontinuous corner, solid line) and $r=2$
  (smooth corner, dashed line).  Fidelity decreases with increasing
  angle $\theta$ in accordance with conventional optical bend-loss.
  Four examples of magnon propagation are shown: (b) $r=0$,
  $\theta=0.15$ (fidelity 0.769), (c) $r=0$, $\theta=0.5$ (fidelity
  0.106), (d) $r=2$, $\theta=0.15$, (fidelity 0.979) (e) $r=2$,
  $\theta=0.5$ (fidelity 0.401).  In all cases, $t_{\rm f}=10$.}
\label{fig:cornersingles}
\end{figure}}

\newcommand{\placeFigBackandForth}{\begin{figure}[t!]
\centering
\includegraphics[width=6cm]{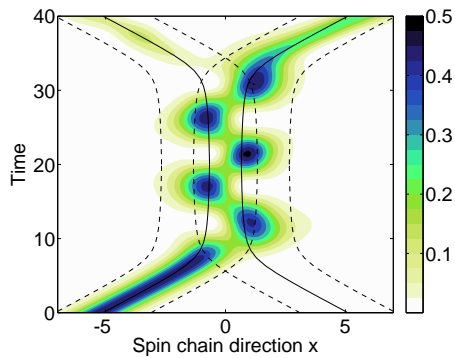}\\
\caption{Magnon evolution through a spin-splitter with  parallel
  component. Oscillatory behavior of the excitation between the
  spin-guides is observed.  Here, $r=0.5$, $d=1.2$, $m=0.3$.}
\label{fig:backandforth}
\end{figure}}

\newcommand{\placeFigBeamsplitters}{\begin{figure}[t!]
\centering
\includegraphics[width=7cm]{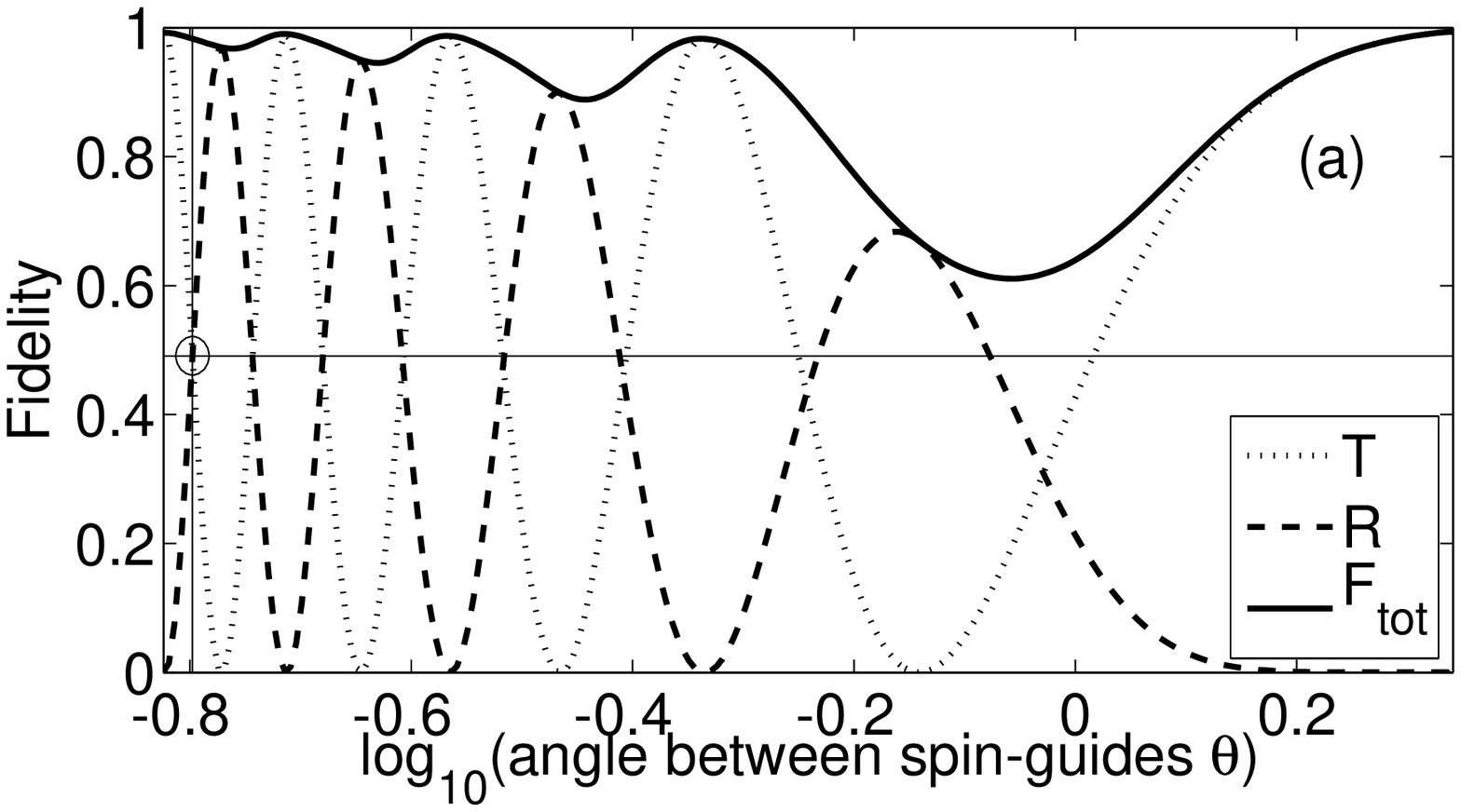}
\includegraphics[width=3.5cm]{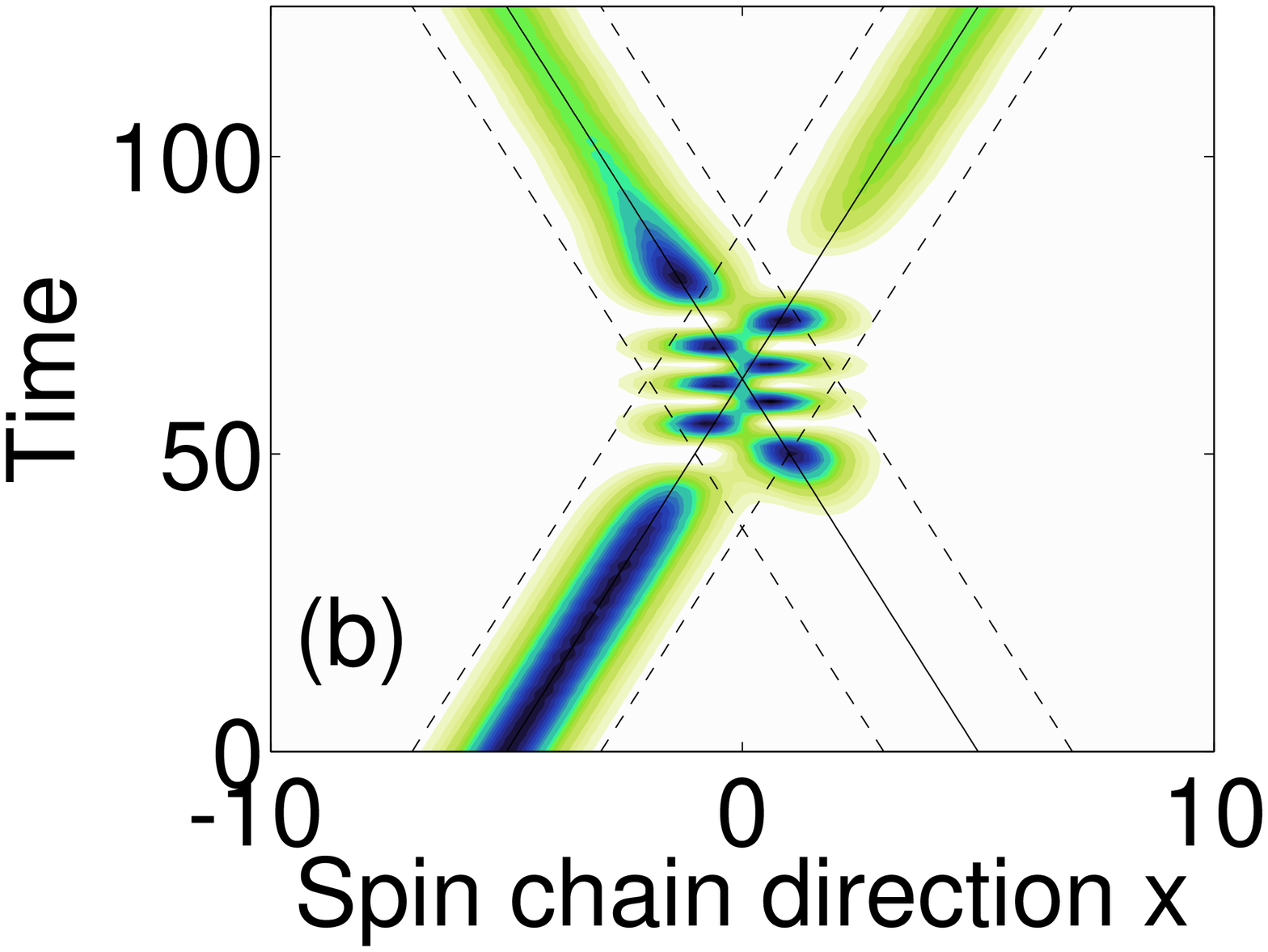} \includegraphics[width=3.5cm]{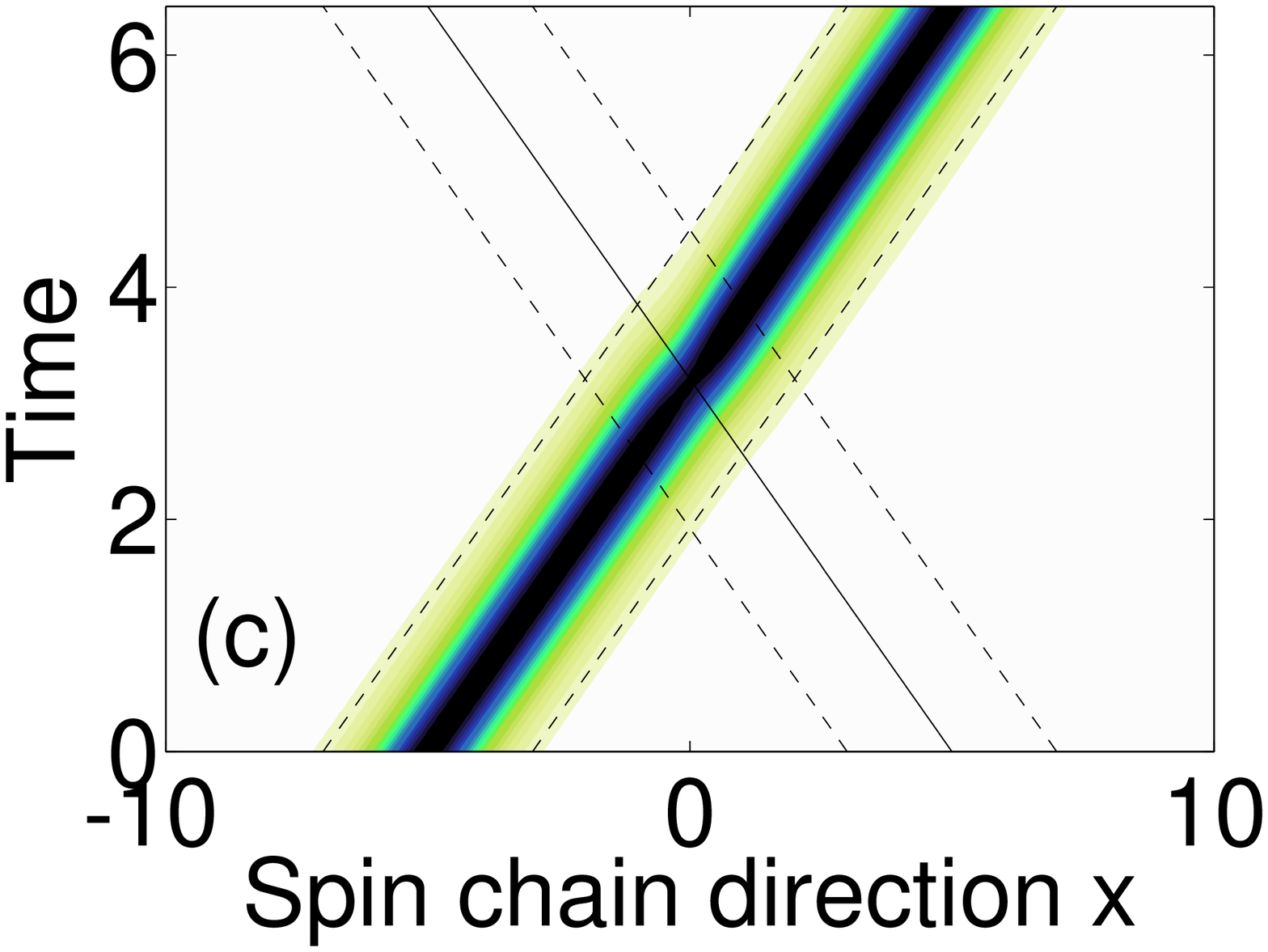}
\caption{(a) Fidelity, and reflection and transmission coefficients
  for an X-junction spin-splitter as a function of spin-guide angle.
  The excitation is initialized in the left-to-right spin-guide and
  then evolved for time $t_{\rm f} = x_l/\tan(\theta/2)$ ($x_l=10$
  fixed).  The reflection and transmission coefficients show
  oscillations as expected from Landau-Zener theory. The
  spin-splitting fidelity is less than unity due to non-adiabaticity,
  although it asymptotes to one in both the small and large angle
  limits.  (b) A 50/50 spin-splitter with $\theta=10^{-0.7976}$
  (corresponding to $R\approx T\approx0.491$, $F_{\rm tot}\approx
  0.982$, indicated by a circle at the intersection of a horizontal
  and a vertical line). (c) A non-adiabatic crossing with
  $\theta=10^{2.2}$.  Note the different time-scales in (b) and (c).}
\label{fig:beamsplitterPhoton}
\end{figure}}

\newcommand{\placePhasePlots}{\begin{figure}[t!]
\centering
\includegraphics[width=4cm]{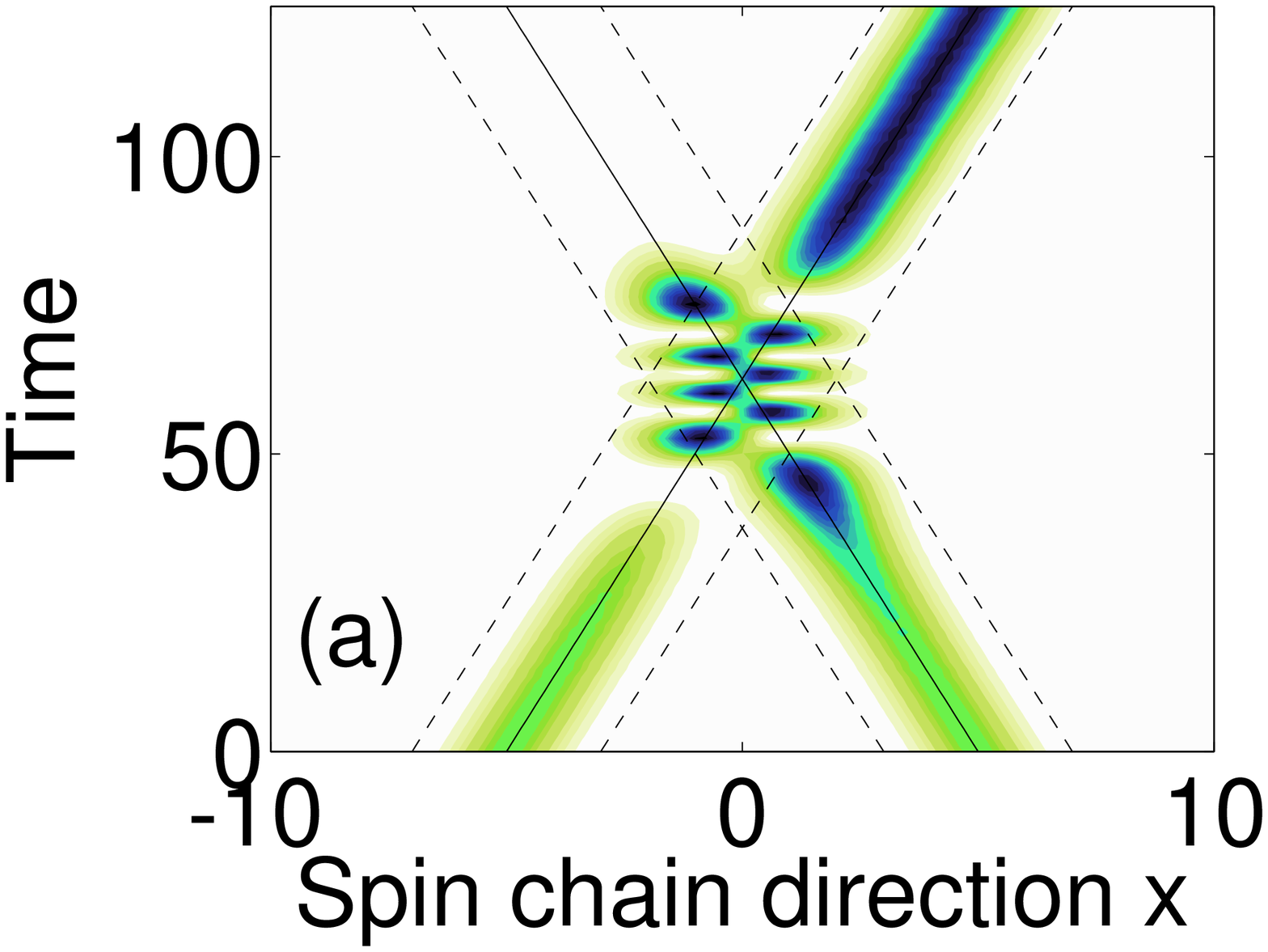}\includegraphics[width=4cm]{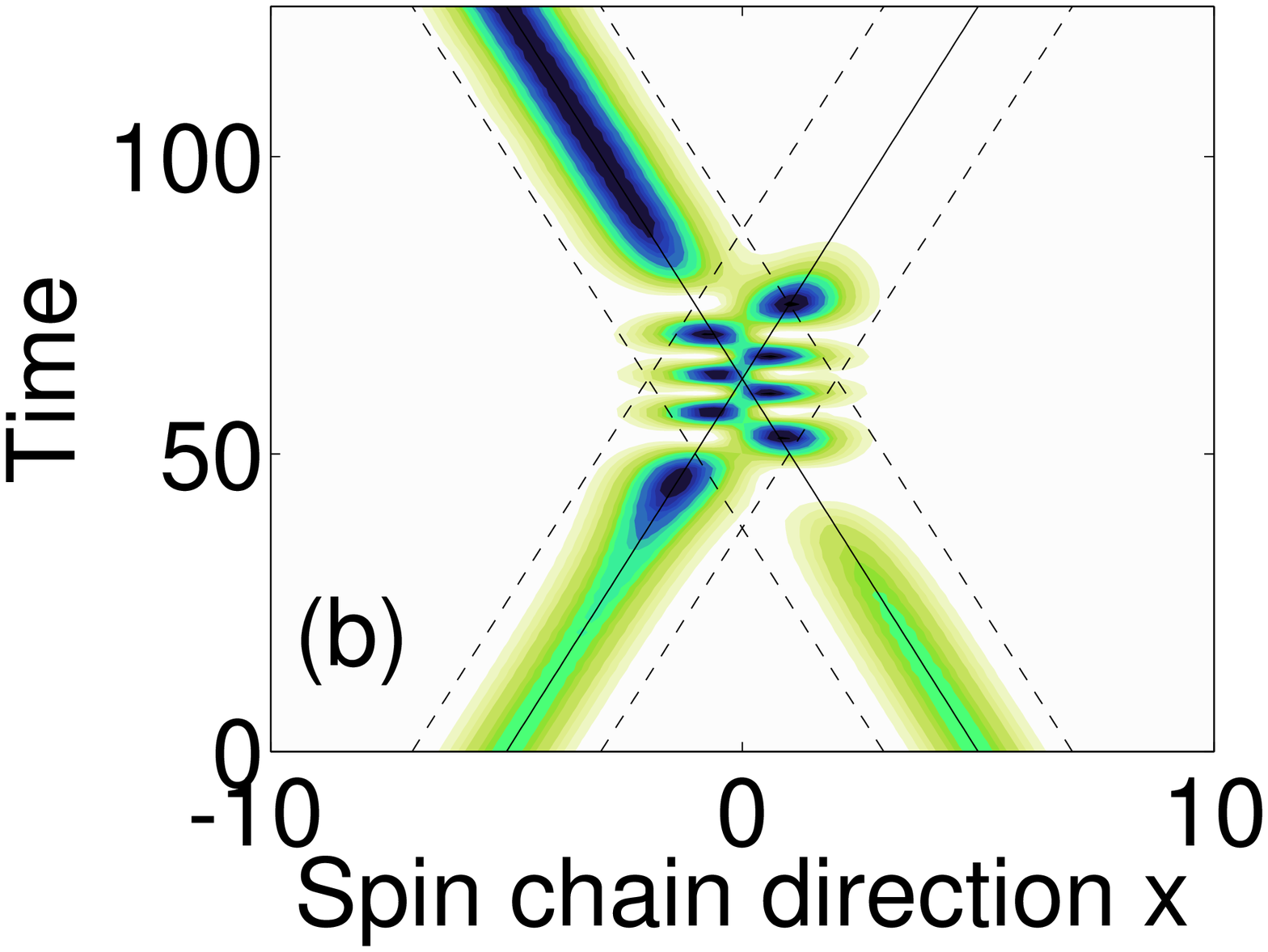}
\includegraphics[height=2cm]{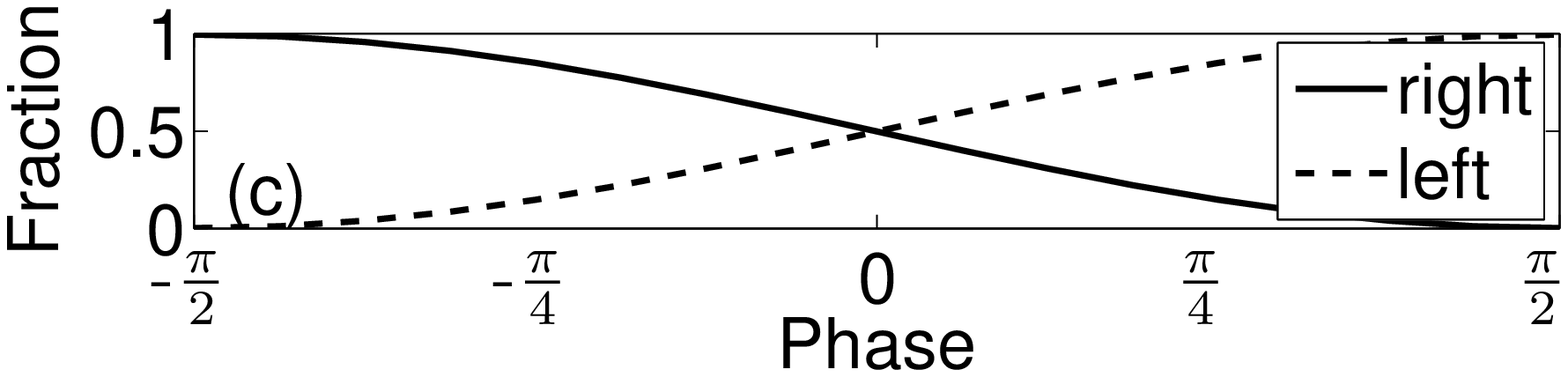}
\caption{(a) [(b)] the evolution when the excitation is initialized as
  a superposition in both spin-guides, the phase between them is
  $\pi/2$ ($-\pi/2$). (c) The fraction of excitation in the left and
  right spin-guides as a function of the phase between the initial
  excitations.}
\label{fig:phasePlots}
\end{figure}}

%%%%%%%%%%%%%%%%%%%%%%%%%%%% TITLE %%%%%%%%%%%%%%%%%%%%%%%%
\title{Spin-guides and spin-splitters: Waveguide analogies in one-dimensional spin chains}

\author{Melissa I. Makin} 
\affiliation{School of Physics, The
  University of Melbourne, Victoria 3010, Australia}

\author{Jared H. Cole}
\affiliation{Applied Physics, School of Applied Sciences, RMIT University, Melbourne 3001, Australia}
\affiliation{Institut f\"ur Theoretische Festk\"orperphysik and DFG-Center for Functional Nanostructures (CFN), Karlsruher Institut f\"ur Technologie, 76128, Germany} 

\author{Charles D. Hill}
\affiliation{School of
  Physics, The University of Melbourne, Victoria 3010, Australia}
\affiliation{ARC Centre for Quantum Computation and
  Communication Technology}

\author{Andrew D. Greentree}
\affiliation{School of
  Physics, The University of Melbourne, Victoria 3010, Australia}

\begin{abstract}
Here we show a direct mapping between waveguide theory and spin chain
transport, opening an alternative approach to quantum information
transport in the solid-state.  By applying temporally varying control
profiles to a spin chain, we design a virtual waveguide or
`spin-guide' to conduct individual spin excitations along defined
space-time trajectories of the chain.  We explicitly show that the
concepts of confinement, adiabatic bend loss and beamsplitting can be
mapped from optical waveguide theory to spin-guides (and hence
`spin-splitters').  Importantly, the spatial scale of applied control
pulses is required to be large compared to the inter-spin spacing, and
thereby allowing the design of scalable control architectures.
\end{abstract}


\maketitle

\placeFigNormalAndUs

The application of quantum information science to technology promises
to make a disruptive change to twenty first century society,
comparable to the computer and telecommunications revolutions of the
twentieth century.  Within this context, there is a pressing need to
develop viable quantum networks.  There have been many proposals to
satisfy this need.  Here we wish to focus on just one implementation
of quantum communication that is ideally suited to solid-state quantum
computing: the one-dimensional spin chain.

The physics of spin chains offers a rich phenomenology. There is a
comprehensive review of the application of spin chains to quantum
information processing due to Bose \cite{ref:boseContemporary}.  In
general, a spin chain is a one-dimensional array of spins that are
closely spaced to facilitate strong spin-spin interactions, perhaps
via dipole-dipole or exchange coupling.  As the inter-spin spacing is
typically on the atomic or near atomic scale, individual
addressability of the spins is either impossible, or unscalable
\cite{ref:copsey, ref:stojanovic, ref:giovannetti, ref:kay}.  As a
consequence of the restriction on local control, many innovative
schemes have been studied to realise spin transport including schemes
with uniform spins and control over just the ends of the chains (see
Refs.~\cite{ref:boseContemporary, ref:stojanovic, ref:giovannetti,
  ref:kay, ref:maloshtan, ref:mogilevtsev}), or with carefully
designed coupling schemes \cite{ref:christandletal, ref:ohshima}.
There has also been related work in transport in coupled cavity
systems \cite{ref:zhou, ref:liao, ref:timedependent}.

Here we outline a distinct alternative to the problem of long-range
quantum information transport inspired by optical waveguides.  We
demonstrate that it is possible to create a virtual waveguide or
`spin-guide' in a one-dimensional spin chain to guide individual spin
excitations, magnons \cite{ref:vleck}, as depicted in
Fig.~\ref{fig:normalAndUs}(a).  An optical waveguide is essentially a
two-dimensional structure, where confinement of an optical mode is
achieved in one dimension by a change in the refractive index of the
medium as a function of space, and the mode propagates in the other
dimension, Fig.~\ref{fig:normalAndUs}(b). Our virtual waveguide uses a
time-varying potential that is controllably swept across the
one-dimensional spin-chain, as depicted schematically in
Fig.~\ref{fig:normalAndUs}(c).  In essence, the two-dimensional
waveguide is replaced by a $1+1$ dimensional spin-guide.  This
approach allows a direct translation of all of the well-known results
from conventional waveguide optics \cite{ref:ladouceur}, and therefore
opens a fundamentally new approach to the manipulation of excitations
in spin chains.  It should be noted that the use of magnons with
Gaussian spatial distributions \cite{ref:osborne}, and adiabatic
following of a locally applied control field \cite{ref:skinner,
  ref:taylor}, have been considered, but we are not aware of any
scheme that has applied the physics of optical waveguiding to coherent
evolution of a solid-state excitation.

To realise the spin-guide, we require a spatially and temporally
varying control field that breaks the degeneracy of the spins in the
chain.  Although the exact system and mechanism for breaking the spin
degeneracy is not essential, for concreteness we consider a
one-dimensional Heisenberg spin chain with a temporally and spatially
varying magnetic field.  Note that although optical waveguides can
usually house many excitations, we are explicitly only considering the
one-excitation subspace, i.e.\ a single magnon.

There is considerable flexibility about the precise choice of applied
magnetic field, and for simplicity we choose a P\"oschl-Teller (PT) potential, for
which numerous analytical results are known \cite{ref:PT,
  ref:rosenmorse, ref:eckart}.  In general any potential that can be
used for optical waveguiding can easily be translated into the
spin-chain model.  By adiabatically varying the \PT potential as a
function of time, the magnon can be guided through a space-time map in
a fashion that is entirely analogous to conventional optical
waveguiding.  The demonstration of this analogy is the central result
of this work.

The Hamiltonian for a system of $N$ spin 1/2 particles with an applied
field is 
\be
\label{eq:discreteH}
H = -J \sum_{n=1}^N \mathbf{S}_n . \mathbf{S}_{n+1} - B(n,t)
S^z_n, 
\ee
where $J$ is the exchange interaction strength, $\mathbf{S}_n$
and $S^z_n$ are operators for the the total spin and the $z$
projection respectively for spin $n$, and $B(n,t)$ is the time-varying
magnetic field applied to spin $n$.

As the control fields are slowly varying across the spin-spin
separation, we can replace the discrete spin chain Hamiltonian
Eq.~(\ref{eq:discreteH}) with its continuum counterpart $\mathcal{H}$
and solve the Schr\"{o}dinger equation,
\be
\label{eq:betabiggerthankappa}
i \frac{\partial}{\partial t} \psi = \mathcal{H}\psi = \left[B(x,t)  - \frac{J}{2} \frac{\partial^2}{\partial x^2} \right]\psi,
\ee
for the evolution as a function of position, $x$.  This is much less
computationally expensive for a large number of spins, yet still
captures all of the essential features of our scheme.  The continuum
limit is important for practical atomic cases, as the spins are
typically separated by one, or  a few, lattice sites, but the
control fields are derived from surface gates and hence are on the
tens of nanometres scale.

We first consider the case of a single spin-guide.  The form of the
\PT potential is $B(x,t) = -B_0 \sech^2[(x-x_0(t))/w]$, where the time
dependence is determined by the moving center of the potential,
$x_0(t)$. For simplicity in what follows, we set $J=1$, $B_0=1$ and
$w=1$.

The magnon state is initialized as the lowest energy eigenstate of the
moving \PT potential, i.e.
\be
\label{eq:initialState}
\psi(x,t=0) = e^{i k x} \sech(x-x_0(0))/\sqrt2,
\ee
where $x_0(0)$ is the centre of the excitation (equivalent to the
center of the spin-guide), the initial momentum $k$ is set to match the initial
velocity of the spin-guide.  Throughout, we solve numerically for $\psi(x,t)$ and display
$|\psi(x,t)|^2$.

\placeFigMoving

To study the effectiveness of the channel, we examine the spin-guide
fidelity. The fidelity is given by the overlap between the initial
wave function $\psi(x,t=0)$, and the excitation at the final time
$t_{\rm f}$ (shifted back to the original location,
$\psi^*[x+x_0(t_f),t=t_{\rm f}]$):
\be
\label{eq:fidelity}
F =\left| \int_{-\infty}^{\infty} \psi(x,t=0)
\psi^*[x+x_0(t_{\rm f}),t=t_{\rm f}]dx \right|^2,
\ee
and $0\leq F\leq 1$.

An important concept with optical fibres is bend-loss, i.e.~the extent
to which an optical fibre can be bent before the mode ceases to be
guided, and is therefore lost.  The equivalent case is accelerating
the magnon by investigating a single spin-guide with a `corner'.
The centre of the spin-guide is given by
\be
\label{eq:ccorner}
x_0(t) = \sqrt{r + \tan^2\theta( t - t_{\rm f}/2)^2} + \tan\theta (t -
t_{\rm f}/2) \ee
where $r$ indicates the sharpness of the corner, $\theta$ is the
angle through which the spin-guide changes direction, and $t_{\rm f}$
is the final time.  The excitation is initially centered at position
$x_0(0)$, with momentum $k = \left.\partial x_0(t)/\partial
t\right|_{t=0}$.

Fig.~\ref{fig:cornersingles}(a) utilises Eq.~(\ref{eq:ccorner}) to
show how the fidelity decreases with increasing angle through which
the spin-guide moves $\theta$.  Two lines are shown: $r=0$
(discontinuous corner) always has lower fidelity than $r=2$ (smooth
corner). As the corner is made more abrupt the fidelity decreases, in
accordance with our intuition from optical bend-loss results.  The
evolution with different examples of $r$ and $\theta$ is shown in
Fig.~\ref{fig:cornersingles}(b)-(d).

\placeFigBackandForth

To complete the connection between spin-guides and waveguides, we turn
our attention to two-port devices, i.e.~we show how to create a
`spin-splitter' by analogy with beamsplitters.  We firstly examine a
spin-splitter with a parallel component, see
Fig.~\ref{fig:backandforth}.  The centre of the potentials of the left
and right spin-guides are given by the piecewise continuous function
\begin{eqnarray}
\label{eq:beamsplitterParallel}
x_{\rm right}(t) &=& -x_{\rm left}(t) =\left\{\begin{array}{ll}
f(x,t) & t < t_{\rm f}/2\\
f(x,t_{\rm f}-t) & t \geq t_{\rm f}/2
\end{array}\right. ,\\
f(x,t)&=&\sqrt{r + \left(4m t + d - x_{\rm l}\right)^2/16}- m t +(d+ x_{\rm l})/4,\nonumber
\end{eqnarray}
where $m$ is the slope and $d$ is the separation between the
parallel components of the spin-guide.  The position of the excitation
is initially in the left spin-guide, i.e. $\psi(x=x_{\rm
  left}(0),t=0)$, and the initial momentum is the slope of the left
spin-guide at time $t=0$, that is $k = \left.\partial x_{\rm
  left}(t)/\partial t \right|_{t=0}$.  In Fig.~\ref{fig:backandforth}
the solid lines show $x_{\rm left}$ and $x_{\rm right}$, and the
dashed lines show $x_{\rm left}\pm2$ and $x_{\rm right}\pm2$, which
can be intuitively thought of as the `edge' of the spin-guides.  The
excitation, after initially starting in the left spin-guide,
oscillates between spin-guides, before leaving primarily through the
right spin-guide.  As expected, the length of the parallel section
compared to the oscillation frequency controls the final output
distribution. This behavior is important to note, as it corresponds to
the small angle limit of the spin-splitter in the next section.

\placeFigBeamsplitters

Coupling between spin-guides is the equivalent of evanescent tunneling
between optical waveguides.  Consider two parallel spin-guides,
separated by distance $d$, with an initial excitation which has zero
momentum, $k=0$.  The oscillation frequency into and out of the left
waveguide is found by expressing the evolution as a two-state problem
and defining an effective Hamiltonian in the basis $\{\psi_{\rm
  left}(x,t),\psi'_{\rm right}(x,t),\ldots\}$, where, $\psi'_{\rm
  right}(x,t)$ is determined by Gram-Schmidt orthonormalization
relative to $\psi_{\rm left}(x,t)$ and $\psi_{\rm right}(x,t)$ which
are given by Eq.~(\ref{eq:initialState}).  The resulting oscillation
frequency is (given by the difference between the eigenvalues of the
effective Hamiltonian),
\be
\Omega(d) =  \frac{{\rm csch}^2d[  {\rm cosh}\,3d + 4 d \,{\rm sinh}\,d-(8d^2+1){\rm cosh}\,d]}{{\rm cosh}\,2d -2d^2 - 1}.
\ee
This function monotonically decreases from $\Omega(0) = 16/15$ to
$\Omega(d) \approx 4 e^{-d}$ for large $d$.

A more practical form of spin-splitter than the style in
Fig.~\ref{fig:backandforth} is an X-junction of spin-guides.  Two
straight spin-guides of length $2x_{\rm l}$ cross at an angle
$\theta$, where $x_{\rm R\rightarrow L}(t) = -x_{\rm L\rightarrow
  R}(t) = x_{\rm l}/2 - \tan(\theta/2) t$.  The initial excitation is
placed in the left-to-right spin-guide, such that $x_0(0) = x_{\rm
  L\rightarrow R}(0)$, with momentum $k_{\rm L\rightarrow
  R}=\tan(\theta/2)$.

The evolution for the X-junction is shown in
Fig.~\ref{fig:beamsplitterPhoton}, for time $t_{\rm
  f}=x_l/\tan(\theta/2)$. The relevant metrics for the spin-splitter
are shown in Fig.~\ref{fig:beamsplitterPhoton}(a) as a function of
spin-guide angle. The reflection and transmission coefficients $R$ and
$T$, as defined by Eq.~(\ref{eq:fidelity}), and the total fidelity is
$F_{\rm tot}=R+T$. The behavior of $R$ and $T$ can be understood with
regard to Landau-Zener theory \cite{ref:landau, ref:zener}.  When
$\theta$ is large, the spin-splitter shows a non-adiabatic crossing,
therefore the reflection (transmission) coefficient approaches zero
(one) for large $\theta$, Fig.~\ref{fig:beamsplitterPhoton}(c).
Conversely, when $\theta$ is small, the spin-guides approach an almost
parallel state.  As such, the excitation behaves similarly to
Fig.~\ref{fig:backandforth}, where the excitation oscillates between
spin-guides and therefore the fidelity depends strongly on $\theta$.
In contrast to conventional Landau-Zener this is not the adiabatic
regime as the spin-guides are forced to cross and the interaction time
increases with decreasing angle so that oscillations are always
observed.

\placePhasePlots

An important spin-splitting ratio is 50/50 ($T=R$), and
Fig.~\ref{fig:beamsplitterPhoton}(a) shows many points where this
ratio is approximately achieved, although with varying fidelity.  An
example, with $\theta=10^{-0.7976}$ is shown in
Fig.~\ref{fig:beamsplitterPhoton}(b), corresponding to the value of
$\theta$ indicated by the circle intersecting the horizontal and
vertical lines in Fig.~\ref{fig:beamsplitterPhoton}(a), when $R\approx
T\approx 0.491$ ($F_{\rm tot}\approx 0.982$). These are slightly less
than 0.5 due to scattering into non-bound modes.  By choosing a
sufficiently small $\theta$, one can generate a 50/50 spin-splitter
with a $T$ and $R$ arbitrarily close to 0.5.

Finally, using our spin-splitter we demonstrate the fundamental
quantum mechanical characteristic of a beamsplitter: the interference
of paths due to their relative phase.  The wave function is
initialized as $\psi(x,t=0) = \frac{1}{\sqrt 2} [ \psi_{\rm
    L\rightarrow R}(x,0)+$ $ e^{i\alpha} \psi_{\rm R\rightarrow
    L}(x,0)]$.  Fig.~\ref{fig:phasePlots}(a) shows the evolution when
$\alpha =\pi/2$ , where the excitation emerges solely from the
left-to-right spin-guide.  Similarly, $\alpha=-\pi/2$ results in the
excitation emerging from the right-to-left spin-guide
[Fig.~\ref{fig:phasePlots}(b)]. The fraction in the left and right
halves of the spin-chain as a function of the phase $\alpha$ is given
in Fig.~\ref{fig:phasePlots}(c).  This phase interference proves the
quantum mechanical nature of the spin-splitter, thereby completing
the analogy between optical waveguides and the behavior of
spin-guides.

We have shown that a collective excitation within a linear spin-chain
can be confined and manipulated using localized field
modulation. Specifically, we see that the space-time behavior of this
one-dimensional excitation mimics effects traditionally observed with
linear optics experiments (in two spatial dimensions).  Using suitably
chosen field modulations in space and time, we can replicate optical
guiding modes, beam-splitting and even phase interference.  This
technique provides a new conceptual framework and method for
controlling spin excitations using field modulation over distances
much greater than the spin-spin separation.  

\section{Acknowledgments}

M.I.M.~acknowledges the support of the David Hay
award. C.D.H.~acknowledges support from the Australian Research
Council Center of Excellence for Quantum Computation and Communication
Technology (Project Number CE110001027).  A.D.G.~acknowledges the
Australian Research Council for financial support (Project
No.~DP0880466).

\bibliographystyle{apsrev} 
\bibliography{../../../thesis/papers}

\end{document}